\documentclass[%
 reprint,
 amsmath,amssymb,
 aps,
 prl,
superscriptaddress
]{revtex4-1}

\usepackage{graphicx}
\usepackage{dcolumn}
\usepackage{bm}

\newcommand{\tsup}[1]{\textsuperscript{#1}}
\newcommand{\psisnj}{\psi_\text{snj}}
\newcommand{\psinj}{\psi_\text{nj}}

\newcommand{\psisnjbgamma}{\psi_\text{snj}^{\scriptscriptstyle\{b,\gamma\}}}

\newcommand{\qop}[1]{\hat{#1}}
\newcommand{\bgammaset}{\scriptscriptstyle\{b,\gamma\}}
\newcommand{\hefour}{\textsuperscript{4}He}

\usepackage{times}
\usepackage[normalem]{ulem}
\usepackage{color}

\begin{document}

\preprint{APS/PREPNUM}

\title{Quantum Phase Transition with a Simple Variational Ansatz}

\author{Y. Lutsyshyn}
\email{yaroslav.lutsyshyn@uni-rostock.de}
\homepage{\\www.physik.uni-rostock.de/qtmps}
\affiliation{%
 Institut f\"ur Physik, Universit\"at Rostock, 18051 Rostock, Germany
}%
\author{G.\ E.\ Astrakharchik}%
\homepage{http://bqmc.upc.edu}
\affiliation{%
Departament de F\'isica i Enginyeria Nuclear, Universitat Polit\`ecnica de Catalunya, Campus Nord B4-B5, E-08034 Barcelona, Spain
}%
\author{C.\ Cazorla}
\affiliation{
School of Materials Science and Engineering, University of New South Wales, Sydney, NSW 2052, Australia
}%
\author{J.\ Boronat}
\homepage{http://bqmc.upc.edu}
\affiliation{%
Departament de F\'isica i Enginyeria Nuclear, Universitat Polit\`ecnica de Catalunya, Campus Nord B4-B5, E-08034 Barcelona, Spain
}

\date{\today}

\begin{abstract}
We study the zero-temperature quantum phase transition between liquid and hcp solid \hefour.
We use the variational method with a simple yet exchange-symmetric and fully explicit wavefunction.
It is found that the optimized wavefunction undergoes spontaneous symmetry breaking and describes the 
quantum solidification of helium at 22 atm. The explicit form of the wavefunction allows to consider 
various contributions to the phase transition. We find that the employed wavefunction 
is an excellent candidate for describing both a first-order quantum phase
transition and the ground state of a Bose solid.
\end{abstract}

\pacs{Valid PACS appear here}
\maketitle



Properties of solid \hefour\ have regained attention due to a host of unexpected 
physics discovered in the past decade 
\cite{%
Chan2004,%
Chan2004-ObservationOfSuperflowInSolidHelium,%
Beamish2007-LowTemperatureShearModulusChangesInSolidHe4AndConnectionToSupersolidity,%
Balibar2010-AnomalousSofteningOfHe4Crystals,%
Chan2013-OverviewOnSolidHe4AndTheIssueOfSupersolidity%
}.
Most of the new features occur close to absolute zero and are believed to
be primarily driven by quantum effects.  
Consequently, the solidification of \hefour\ came under renewed scrutiny.
The role of quantum statistics in the transition location has been recently
revisited in
Ref.~\cite{Prokofev2012-RoleOfBoseStatisticsInCrystallizationAndQuantumJamming}.
At small but non-zero temperatures, indistinguishability of particles
destabilizes the quantum solid.  Distinguishable particles, on the other hand,
would solidify even at low pressures, with the phase diagram reminiscent of
the Pomeranchuk effect
\cite{Pomeranchuk1950-TheoryOfLiquidHe3,Richardson1997-ThePomeranchukEffect}.
The feature was dubbed in
\cite{Prokofev2012-RoleOfBoseStatisticsInCrystallizationAndQuantumJamming}
as thermocrystallization.   A similar effect was seen numerically for the
Wigner-crystallization of a 2D Coulomb system
\cite{Ceperley2009-HexaticAndMesoscopicPhasesInA2DQuantumCoulombSystem}.
The solidification of \hefour\ at zero temperature was revisited in
Ref.~\cite{Galli2013-DensityFunctionalTheoryAndBoseStatisticsForTheFreezingOfSuperfluidHe4}
with the density functional theory (DFT). Results were improved comparing with
previous DFT studies. 

In this Letter, we  show  that the quantum solidification of \hefour\ can be considered variationally, 
with a single explicit wavefunction which selects the phase
through optimization of the thermodynamic potential. 
Quite surprisingly, we find that the phase transition is predicted properly, 
given the relative simplicity of the  wavefunction. While the variational treatment is used for 
quantum phase transitions at the mean-field level \cite{Sachdev2001QuantumPhaseTransitions}, it is relatively uncommon that a (discontinuous) transition can be described with a microscopic wavefunction.

At zero temperature, the phases of \hefour\ can be studied in an
essentially exact form  with a family of projector methods, including Green's-function Monte Carlo \cite{Chester1979-PropertiesOfLiquidAndSolidHeFour, Chester1981-ModernPotentialsAndThePropertiesOfCondensedHeFour},
diffusion Monte Carlo
\cite{Boronat2005-QuantumMonteCarloSimulationOfOverpressurizedLiquidHe4},
and path integral ground state Monte Carlo \cite{Schmidt200-APathIntegralGroundStateMethod}.
These methods properly describe the phase transition in helium, and can provide insight 
on the nature of its ground state
\cite{Reatto1987-MaximumOverlapJastrowWaveFunctionForLiquidHeFour,Waintal2007-VariationalWaveFunctionsAndTheirOverlapWithTheGroundState}.

Variational calculations with shadow-type wavefunctions (SWF) %
\cite{Vitiello1988-VariationalCalculationsForSolidAndLiquidHe4WithAShadowWaveFunction}
provide accurate results both for the liquid and solid phases of
\hefour
\cite{Vitiello1988-VariationalCalculationsForSolidAndLiquidHe4WithAShadowWaveFunction,Reatto1998-VariationalTheoryOfBulkHeFourwithShadowWaveFunctions}, 
and describe the transition
\cite{Reatto1995-HomogeneousNucleationOfCrystallineOrderInSuperdenseLiquidHeFour,Chester2000-MonteCarloStudiesOfTwoDimensionalPhasesOfHeliumUsingAShadowWaveFunction,Vitiello1994-AnImprovedShadowWavefunctionForBulkHeFour} and coexistence \cite{Reatto1994-QuantumTheoryOfSolidLiquidCoexistenceAndInterfaceInHeFour} between the two phases.
The SWF can be seen as representing a single step of a projection calculation 
\cite{Vitiello1988-VariationalCalculationsForSolidAndLiquidHe4WithAShadowWaveFunction}.
The projection is carried out by performing the numerical integration of the shadow degrees of freedom.
In this sense, the SWF is not fully explicit, as one cannot write down the result of such an integration.
We consider SWF calculations as a class of their own, in between the exact projection methods and 
the simple and fully explicit 
wavefunction used here.
 
Highly effective wavefunctions have been developed over the years for liquid and (nonsymmetric) solid helium. 
Accurate multi-parameter two- and three- body terms 
\cite{Pandharipande1973-VariationalMethodForDenseSystems,%
Vitiello1992-OptimizationOfHeWaveFunctionsForTheLiquidAndSolidPhases,%
Vitiello1999-VariationalMethodsForHeUsingAModernHeHePotential} 
result in energies that are nearly exact
\cite{Vitiello1999-VariationalMethodsForHeUsingAModernHeHePotential}.
However, efficient one-body (lattice) terms that are also exchange-symmetric have not been reported.

\begin{figure*}[tb]
\includegraphics[angle=0,width=0.32\textwidth]{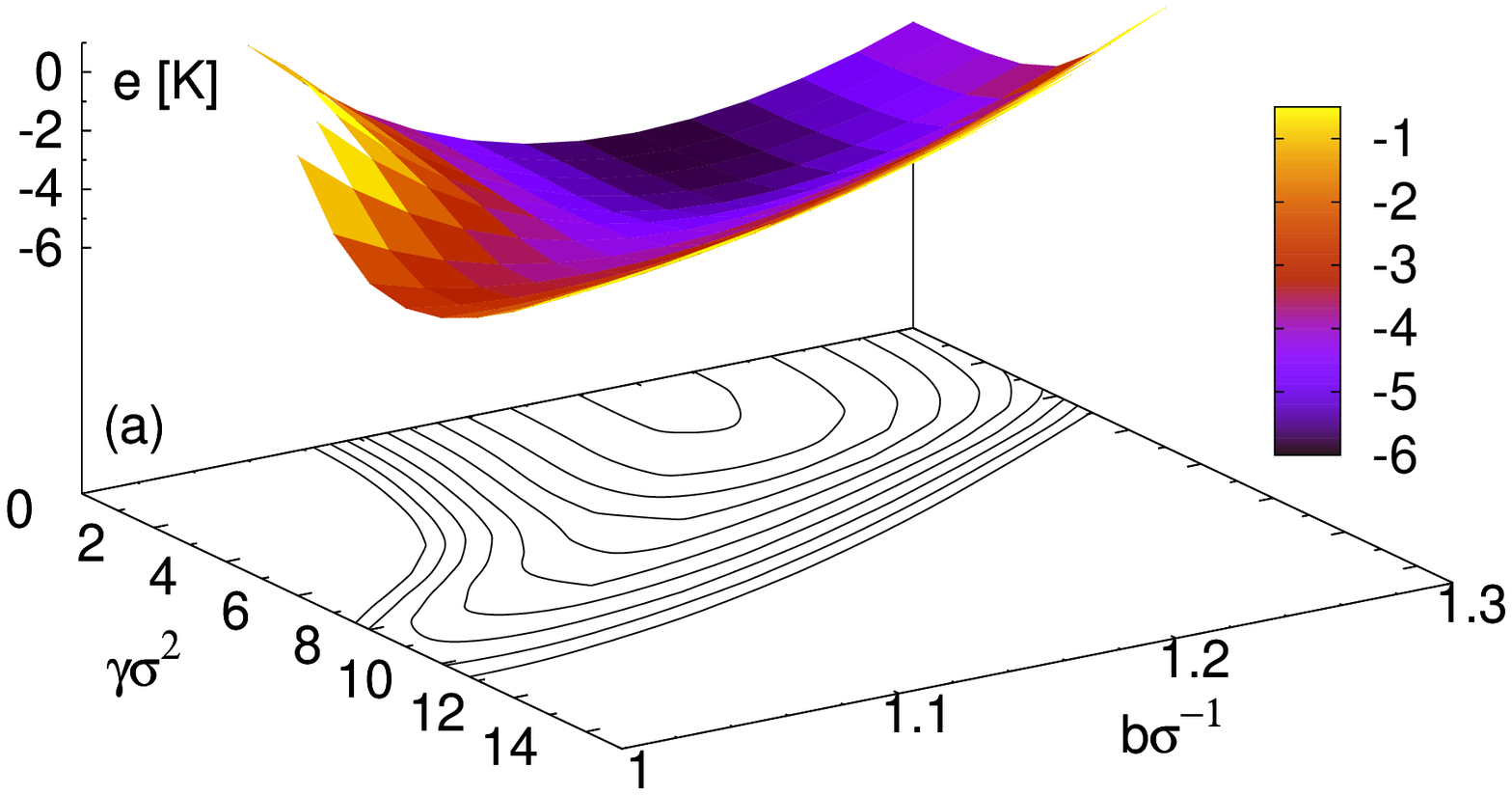}\hspace*{3px}
\includegraphics[angle=0,width=0.32\textwidth]{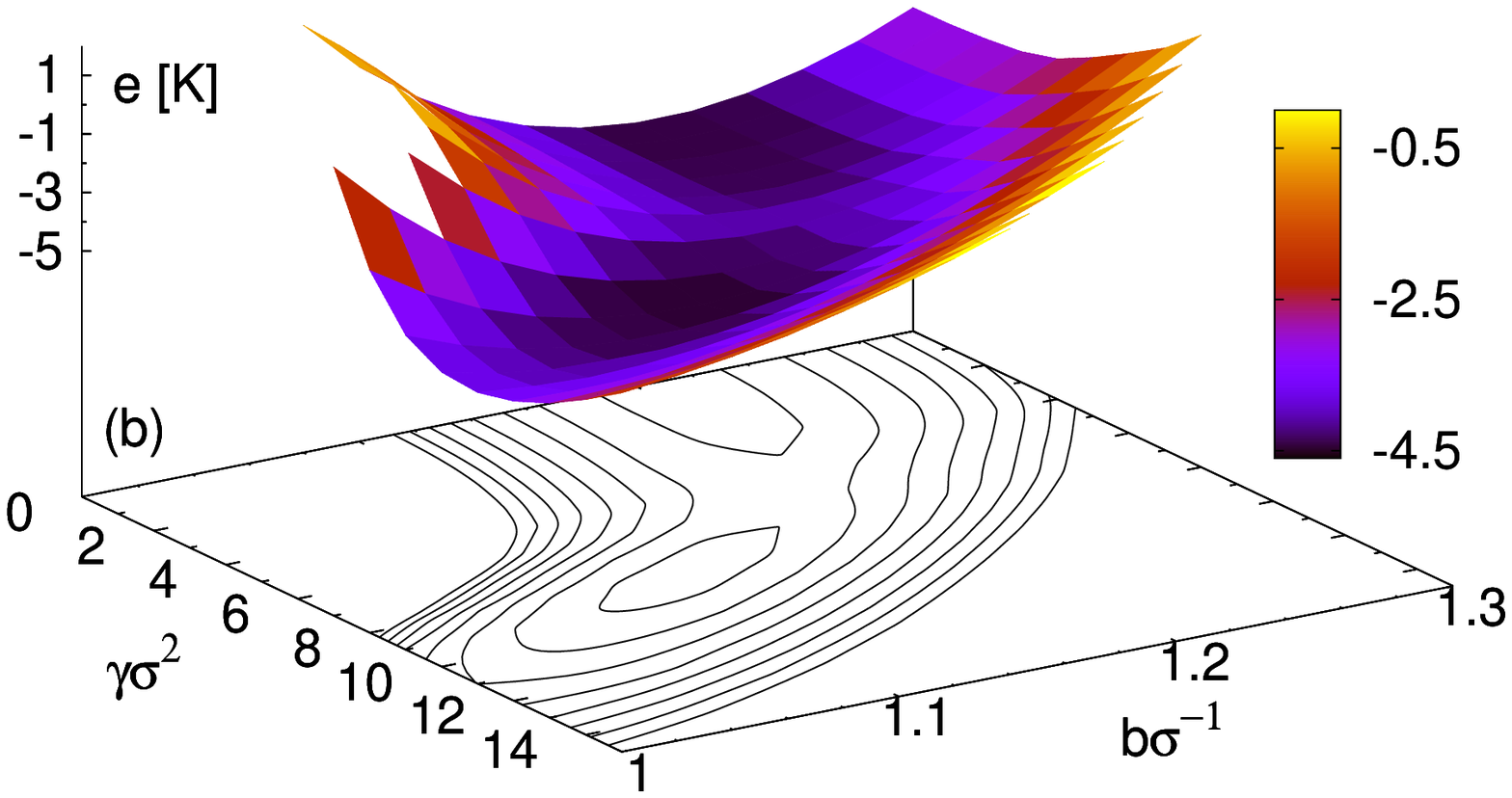}\hspace*{3px}
\includegraphics[angle=0,width=0.32\textwidth]{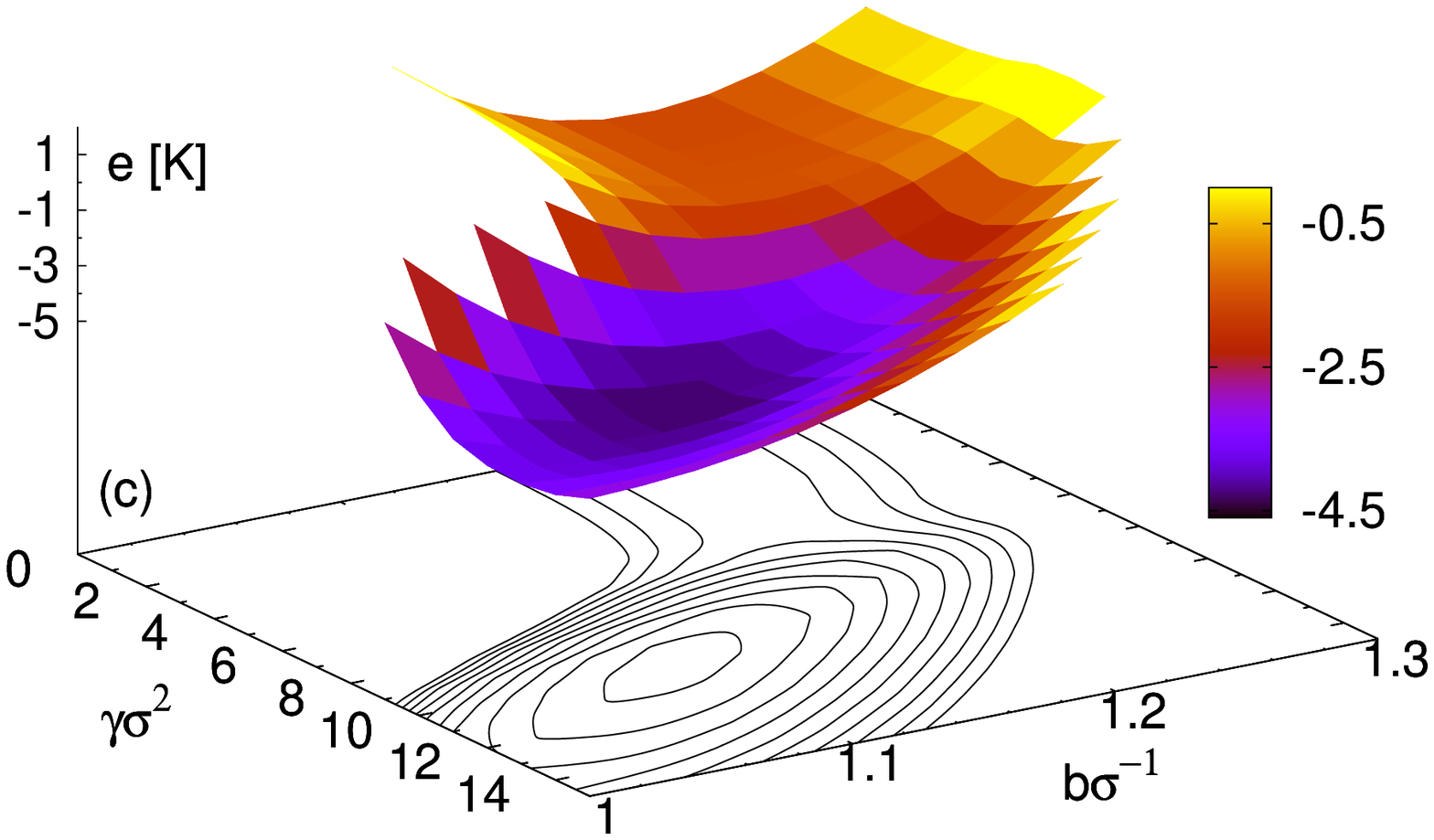}
\caption{\label{fig:Surfaces}
(Color online) Energy per particle obtained with the two-parameter symmetric wavefunction $\psisnjbgamma$
given by Eqs.~(\ref{eq:psisnj}-\ref{eq:locfactor}),
as a function of variational parameters 
$b$ and $\gamma$, for three different densities.
Parameters are shown in terms of $\sigma=2.556$\AA.
Simulation used $N=N_\text{s}=900$ particles.
(a): density $\rho=22.2$~nm\tsup{$-3$} displays a single minimum, at $\gamma=0$, which corresponds to a liquid phase;
(b): intermediate density $\rho=25.8$~nm\tsup{$-3$} is in the 
liquid-solid coexistence region, and has two separate local minima;
(c): density $\rho=29.3$~nm\tsup{$-3$} displays only one 
minimum, corresponding to the solid phase. Contours are separated by 0.4~K.
}
\end{figure*}

The wavefunction that we consider here was proposed specifically for solid \hefour\ \cite{Boronat2009NJP}.
It has been since then used extensively for importance sampling in projector Monte Carlo methods 
\cite{%
Boronat2013-PossibleSuperfluidityOfMolecularHydrogenInATwoDimensionalCrystalPhaseOfSodium,%
Boronat2013-ElasticConstantsOfIncommensurateSolidHe4FromDiffusionMonteCarloSimulations,%
Lutsyshyn2010-PropertiesOfVacancyFormationInHcp4HeCrystalsAtZeroTemperatureAndFixedPressure%
}.  
This wavefunction is a product of the Jastrow term, which accounts 
for the pair correlations, and a cleverly 
symmetrized  Nosanow-like term. 
The wavefunction has the the form
\begin{equation}
\psisnj= 
\left(\prod_{i<j}^{N_\text{p}} f\left(\left|\bm{r}_i-\bm{r}_j\right|\right) \right)
\left( \prod_{k}^{N_{\text{s}  \makebox[0cm]{\phantom{$\scriptscriptstyle p$}}     }} 
\sum_i^{N_\text{p}} g\left(\left|\bm{r}_i-\bm{l}_k\right|\right)\right),
\label{eq:psisnj}
\end{equation}
where $N_\text{p}$ is the number of atoms, and $N_s$ is the number 
of lattice sites, located at $\bm{l}_k$. Position
of the $i$\tsup{th} particle is labeled $\bm{r}_i$.
Suitable for our interest in the thermodynamic limit, $\psisnj$ has the translational invariance broken by the lattice site locations $\bm{l}_k$.
Notice that the second, product-sum term in Eq.~(\ref{eq:psisnj}), is \emph{not} a permanent, and the computational cost of $\psisnj$ scales only as the square of the number of particles.
Pair correlation factors $f(r_{ij})$ can be taken with the pseudopotential either in the McMillan form \cite{McMillan,SchiffVerlet-1967}
\begin{equation}
f(r)=\exp\left[-\frac12 \left(\frac{b}{r}\right)^5\right],
\label{eq:svfactor}
\end{equation}
or in a more involved form with mid-range correlations, as detailed below.
Atoms are localized to the lattice sites with factors $g(r)$. 
We use the Gaussian form 
\begin{equation}
g(r)=\exp\left[ -\frac12 \gamma r^2 \right],
\label{eq:locfactor}
\end{equation}
with parameter $\gamma$ describing the strength of the site localization.

To better understand the structure of $\psisnj$, we can write the wavefunction in the form
\begin{equation}
\psisnj= 
\prod_{i<j}^{N_\text{p}} f(r_{ij})
\prod_{k}^{N_{\text{s}  \makebox[0cm]{\phantom{$\scriptscriptstyle p$}}     }} 
S_k,
\label{eq:psisnjwithsk}
\end{equation}
with the site-sums $S_k(\bm{r}_1,\dots,\bm{r}_{N_\text{p}})$ given by
\[
S_k=\sum_i^{N_\text{p}} g\left(\left|\bm{r}_i-\bm{l}_k\right|\right).
\]
While each sum $S_k$ depends on the coordinates of all particles,
it does not contain interparticle distances. One can view 
them as a generalized form of one-body correlation factors, in the formal sense that
\mbox{$\bm{\nabla}_{i\ne j}\cdot\bm{\nabla}_{j} S_k = 0$}.
In this view, Eq.~(\ref{eq:psisnjwithsk}) consists of the one- and two-body terms of 
the general Feenberg form for the trial wavefunction \cite{Feenberg1974-GroundStateOfAnInteractingBosonSystem,Campbell1977-EnergyAndStructureOfTheGroundStateOfLiquidHeFour}.
Equation~(\ref{eq:psisnjwithsk}) also emphasizes the flexibility of $\psisnj$. 
The number of sites does not need to be equal to the
number of particles. One may confine atoms to given regions of the lattice by including
these atoms only in some of the sums $S_k$.
Limiting each sum to only one atom recovers the original Nosanow-Jastrow wavefunction \cite{Nosanow1964},
\begin{equation}
\psinj=\left(\prod_{i<j}^{N_{\text{p}}} f\left(\left|\bm{r}_i-\bm{r}_j\right|\right) \right)
\left( \prod_{k}^{N_{\text{s}  \makebox[0cm]{\phantom{$\scriptscriptstyle p$}}     }} 
 g\left(\left|\bm{r}_k-\bm{l}_k\right|\right)\right), \, N_\text{s}=N_\text{p}.
\label{eq:psinj}
\end{equation}
The Nosanow-Jastrow wavefunction $\psinj$ yields good variational energy and has long been used to describe solid \hefour.
Unfortunately, $\psinj$ is not exchange symmetric %
\footnote{One can, in principle, argue that the overlap of any non-symmetrical wavefunction and the true ground state 
of \hefour\ scales with the number of particles $N$ as $(N!)^{-1}$.}.
The one-body term imposes a heavy penalty for removing an atom away from its ``parent'' site.
A straightforward symmetrization of $\psinj$ yields poor results 
\cite{Ceperley1978-MonteCarloStudyOfTheGroundStateOfBosonsInteractingWithYukawaPotentials},
or otherwise results in computationally prohibitive wavefunctions.

The lattice structure $\bm{l}_k$, which enters through the site-localization terms $g(\cdot)$,
can in principle be seen as a parameter to the wavefunction.
In this case one may optimize between different lattice symmetries, or optimize individual site positions. 
On the other hand, lattice can be seen as an input to the problem. Here we follow the latter path,
since we aim to study the experimentally known zero-temperature solid phase of \hefour. Thus $\bm{l}_k$ are located
on a geometrically ideal hcp lattice.

\begin{figure}[tb]
\includegraphics[angle=0,width=\columnwidth]{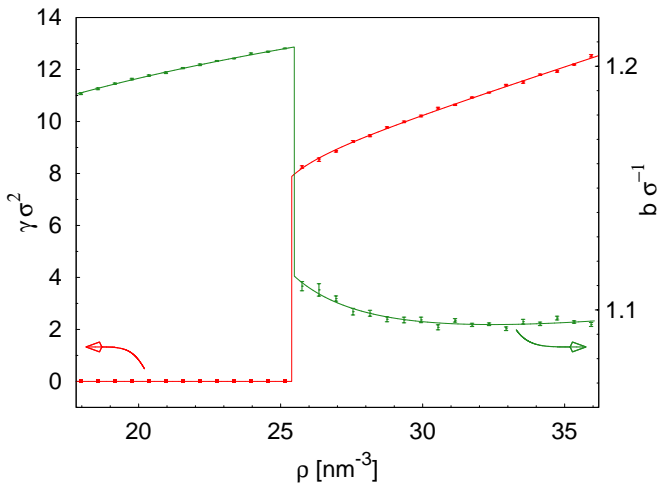}
\caption{ \label{fig:Parameters}
(Color online) Optimized parameters of the two-parameter 
symmetric wavefunction given by Eqs.~(\ref{eq:psisnj}--\ref{eq:locfactor}),
as a function of density.
Parameters are shown in units of $\sigma=2.556$\AA.
Left vertical axis corresponds to the value of the site-localization parameter $\gamma$, while the right axis shows parameter $b$. 
}
\end{figure}

We begin with the variational energy optimization of the two-parameter 
trial wavefunction $\psisnjbgamma$ given 
by Eqs.~(\ref{eq:psisnj}--\ref{eq:locfactor}). 
The energy is given by 
\begin{equation}
E(b,\gamma)=\left\langle \psisnjbgamma \left|\qop{H}\right| \psisnjbgamma \right\rangle
\left/
\left\langle \psisnjbgamma \left| \psisnjbgamma \right\rangle  \right. \right. ,
\label{eq:energyestimator}
\end{equation}
with many-body Hamiltonian 
\[
\qop{H} = -\frac{\hbar^2}{2m_{\scriptscriptstyle\text{He4}}} \sum_{i=1}^{N_\text{p}}{\nabla^2_i} + \sum_{i\ne j}V(r_{ij}),
\]
using the HFD--B(HE) pairwise interaction potential proposed by Aziz et al.~\cite{AzizII}.
The multidimensional integral 
implied by Eq.~(\ref{eq:energyestimator})
was evaluated with a Metropolis Monte Carlo scheme
\cite{Lutsyshyn2014-FastQuantumMonteCarloOnAGPU,GPUFootnote}.
We performed a direct minimization
on a grid of $b$ and $\gamma$ values, for a range of densities.
Three characteristic examples of the energy surface are shown in Fig.~\ref{fig:Surfaces}.
At low densities, there is a single minimum with $\gamma=0$. The wavefunction with $\gamma=0$ reduces to the translationally invariant Jastrow product and corresponds to a liquid phase.
At intermediate densities, an additional local minimum appears at nonzero values of $\gamma$, corresponding to a state with broken translational symmetry. 
This minimum corresponds to a crystalline phase, as was verified from
the scaling with $N_\text{p}$
of the static structure function.
With further increase in density, this second $\gamma\ne 0$ minimum lowers in energy and eventually ``overtakes'' the
liquid $\gamma=0$ minimum. Thus the solid phase becomes preferred variationally, and the optimized system loses translational symmetry.
With the densities increased  further still, the liquid minimum disappears.
The optimal values of parameters $b$ and $\gamma$, shown in Fig.~\ref{fig:Parameters} as a function of density,
display a clear transition between the solid and liquid states.
As our simulated system is finite, the sharpness of this transition is 
in fact a remarkable occurrence 
\cite{Vicari2014-FiniteSizeScalingAtFirstOrderQuantumTransitions,%
Desmond2014-NonstandardFiniteSizeScalingAtFirstOrderPhaseTransitions,%
Duan2011-SharpPhaseTransitionsInASmallFrustratedNetworkOfTrappedIonSpins}.
Despite effort, we were not able to detect any smooth rollover between the phases.
Technically, the two minima in the energy surface, as shown in Fig.~\ref{fig:Surfaces}~(b),
are always distinct. 
We attribute this effect to the fact that the thermodynamic limit 
is accessible to the finite system
through the provided lattice $\bm{l}_k$.

Full thermodynamic analysis requires
minimization of the Gibbs free energy,
which at $T=0$ equals enthalpy, $G=E+PV$.
Enthalpy has to be extracted from the
equations of state $E^{\bgammaset}(\rho)$ computed for each possible set 
of parameters $\{b,\gamma\}$.
The pressure can be computed via
\[
P^{\bgammaset}(\rho)=\rho^2 \frac{\partial E^{\bgammaset}/N}{\partial \rho}.
\]
We solve the above equation for $\rho^{\bgammaset}(P)$ and 
find the Gibbs energy for each set of parameters,
\begin{equation*}\
G^{\bgammaset}(P)=E+PV=E(\rho^{\bgammaset}(P))+P N/\rho^{\bgammaset}.
\end{equation*}
Next, we minimize $G^{\bgammaset}(P)$
with respect to variational parameters,
\[
G(P)=\min_{\bgammaset} G^{\bgammaset}(P).
\]
It is possible to show that only the parameters
which minimize energy at \emph{some} density
will also minimize free energy at any pressure.
The parameters which minimize the free energy
at a given pressure also provide the density and energy at that pressure.
While this  method is relatively straightforward, such analysis has not been reported in the past, presumably 
because of the large underlying computational costs.

\begin{figure}[tb]
\includegraphics[angle=0,width=\columnwidth]{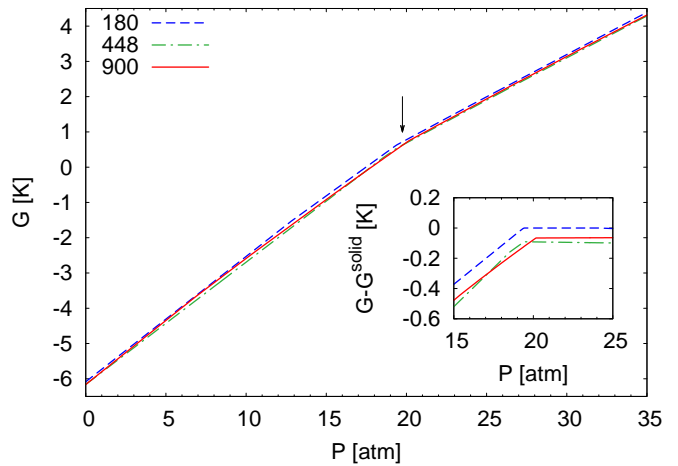}
\caption{\label{fig:Gibbs}
(Color online) Gibbs free energy, per particle, of the optimized state of the two-parameter 
symmetric wavefunction given by Eqs.~(\ref{eq:psisnj}--\ref{eq:locfactor}).
Arrow indicates the location of the phase transition.
The inset shows the free energies with subtracted linear fit to the solid phase of the 180-particles system.
}
\end{figure}

\begin{figure}[t]
\includegraphics[angle=0,width=\columnwidth]{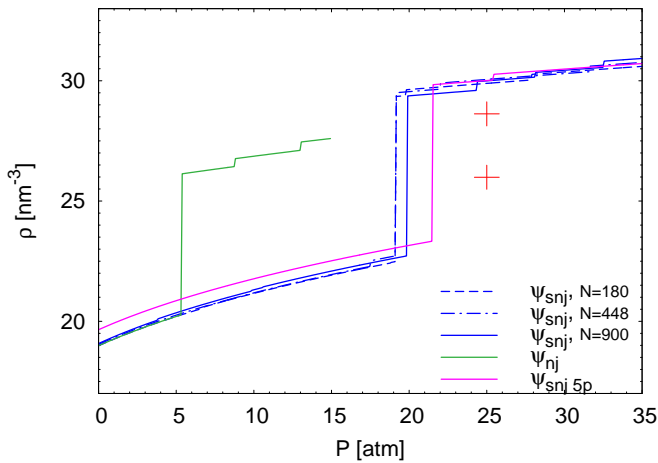}
\caption{   \label{fig:Density}
(Color online) Optimized density 
as a function of pressure. Blue lines show
density for the two-parameter symmetric wavefunction $\psisnj$ for 
varying number of particles, $N=180, 448, \text{and } 900$.
The size effects in the transition location are below the statistical error, which is about 1~atm.
Purple line shows density for five-parameter symmetric 
wavefunction with improved pair-correlation factor~(\ref{eq:reattofactor}), with transition at 22 bar.
Green line shows density for the traditional, non-symmetric Nosanow-Jastrow wavefunction 
of Eq.~(\ref{eq:psinj}), with transition at 5 bar. Red cross marks mark experimental
melting and freezing densities \cite{Ouboter1987,Edwards1965}.
The piecewise appearance of data comes from different equations 
of state $\rho(P)$ selected by the varying pressure-optimized variational parameters; 
size of the steps is thus indicative of the statistical error.
}
\end{figure}

We carried out the minimization procedure outlined above for a range of densities.
The Gibbs free energy, shown in Fig.~\ref{fig:Gibbs} as a function of pressure,
exhibits a kink characteristic of a first-order phase transition.
Corresponding to the weakness of this transition, the kink is subtle yet well-defined.
We performed a linear fit to the free energy of the solid phase near the transition,
and subtracted this fit from the free energies. The result, which emphasizes the transition, is plotted 
in the inset to Fig.~\ref{fig:Gibbs}.
Somewhat unexpectedly, the transition occurs at a pressure of 20~atm, close to the correct value of 25~atm.
This is especially surprising given the simplicity of our two-parameter wavefunction.
At the transition pressure, optimized density jumps from the lower density of freezing $\rho_f$ to the higher density of melting $\rho_m$.
At zero temperature, the density discontinuity also provides information about the latent heat $\Delta E$, since
$
\Delta E = P \left(1/{\rho_f}-1/{\rho_m}\right).
$
Optimized density $\rho(P)$ is plotted in Fig~\ref{fig:Density},
along with the experimental values \cite{Ouboter1987,Edwards1965}. 
Results are shown for several particle numbers, from
$N_\text{p}=180$ to  $N_\text{p}=900$.
As can  be seen, the size effects are moderate and the extrapolated 
transition pressure for the two-parameter wavefunction
is close to 20 atm.

The two-body factors of Eq.~(\ref{eq:svfactor}) account for the short-range behavior of the interaction potential. 
More accurate two-body factors can be obtained by including
mid-range correlations, as  in the form proposed in 
Ref.~\cite{Reatto1979-SpatialCorrelationsAndElementaryExcitationsInManyBodySystems},
\begin{equation}
f(r)=\exp\left[-\frac12 \left(\frac{b}{r}\right)^5  + \frac12 s \exp\left( \frac{r-\lambda}{ w } \right)^2 \right].
\label{eq:reattofactor}
\end{equation}
The resulting five-parameter symmetric wavefunction given by Eqs.~(\ref{eq:psisnj}), (\ref{eq:locfactor}), and (\ref{eq:reattofactor})
was optimized and analyzed as described above. 
For non-zero optimal $\gamma$,
the mid-range correlation factor optimizes away, i.e., \ $s=0$.
That is, the solid phase does not benefit from such correlations.
The results for the optimal density at each pressure are shown in Fig.~\ref{fig:Density}. 
As can be seen, the transition location has improved, and the zero-pressure liquid density has increased, which is closer to the experiment.
We also carried the optimization with the two-body factors that included the low-energy phonon contribution along Ref.~\cite{Reatto1970-GroundStateOfLiquidHeFour}. 
Such terms had little influence on the transition location.
We conclude that the main source of discrepancy in the pressure of the phase transition stems from
deficiencies in the two-body factors and the absence of three-body
correlation factors in the liquid phase.

The optimization of the thermodynamic potential was also carried out for 
the two-parameter (unsymmetrized) Nosanow-Jastrow wavefunction $\psinj$ given by Eqs.~(\ref{eq:svfactor},\ref{eq:locfactor},\ref{eq:psinj}).
The optimization results are included in Fig.\ref{fig:Density}.
Solidification for $\psinj$ occurs already at 5~bar, a dramatic five-fold departure from the correct location of the transition.
In fact, the freezing density $\rho_f$ for $\psinj$ lies below
the correct experimental equilibrium density of helium liquid at vapor pressure.
(The equilibrium density is underestimated by all wavefunctions by as much as 15\%.)
Thus non-symmetric Nosanow-Jastrow wavefunction misses the transition by a wide margin. 
It should be noted that $\psinj$ in fact provides lower energies for the solid than the symmetric $\psisnj$, by up to 1~K.
Strict site localization makes $\psinj$ insensitive to the deficiencies in the two-body factors.
The symmetrical $\psisnj$ allows for virtual interstitials and is more sensitive to the form of two-body factor $f(r)$.
As the balance between phases amounts to the difference between free energies, to some extend a cancellation occurs, improving
the location of the transition for $\psisnj$.

To summarize, we studied at the variational level the quantum phase transition between superfluid  and  hcp solid \hefour.
The transition properties were determined by optimizing the Gibbs free energy.
We used a wavefunction which describes a quantum solid with broken translational 
symmetry but that is, at the same time, exchange-symmetric.
Below the melting pressure, the optimized wavefunction reduces to a translationally 
symmetric Jastrow function describing a liquid.
Given the simplicity of the wavefunction, it is remarkable that the  
transition is found at a pressure that is only  
three to five atm away from the correct experimental value. 
We attribute the discrepancy to the quality of the pair-correlation terms
 and the lack of three-body correlations in the liquid phase.
These findings strongly support the form given by Eq.~(\ref{eq:psisnj}) as a 
suitable symmetric wavefunction for describing both a first-order quantum
phase transition and a quantum Bose solid.


Authors thank the Barcelona Supercomputing Center 
(The Spanish National Supercomputing Center -- Centro Nacional de Supercomputaci\'on)
for the provided computational facilities. We acknowledge partial financial support from 
the DGI (Spain) Grant
No.~FIS2011-25275  and Generalitat de Catalunya Grant No.~2009SGR-1003. 
G.~E.~A.\ acknowledges support from the Spanish MEC through the Ramon y Cajal
fellowship program.



%

\end{document}